\documentclass[prl,showpacs,twocolumn,superscriptaddress]{revtex4}

\usepackage{graphicx, dcolumn, bm}
\usepackage{textcomp}
\usepackage{natbib}
\usepackage{amsmath}
\usepackage{amssymb}

\begin{document}
\title{Direct bandgap silicon quantum dots achieved via electronegative capping}
\author{A. N. Poddubny}\email{poddubny@coherent.ioffe.ru}
\affiliation{Ioffe Physical-Technical Institute of the Russian Academy of Sciences, 26 Politekhnicheskaya st., St. Petersburg 194021, Russia}
\author{K.~Dohnalov\'a}
\affiliation{Van der Waals-Zeeman Institute, University of Amsterdam, Science Park 904, NL-1098 XH Amsterdam, The Netherlands}
\date{\today}

\begin{abstract}
We propose a novel concept of achieving silicon quantum dots with radiative rates enhanced by more than two orders of magnitude up to the values characteristic for direct band gap semiconductors. Our tight-binding simulations show how the surface engineering can dramatically change the density of confined electrons in real- and $k$-space and give rise to the new conduction band levels in $\Gamma$-valley, thus promoting the direct radiative transitions. The effect may be realized by covering the silicon dots with covalently bonded electronegative ligands, such as alkyl or teflon chains and/or by embedding in highly electronegative medium.
\end{abstract}
\maketitle

Light emission from silicon based materials is of great interest for monolithic integration of optoelectronics and photonics with microelectronics \cite{Priolo14, Naturelaserreview, Dohnalova14}. Indirect bandgap bulk silicon is not suitable for this purpose and the most promising platform towards efficient and tunable emission are silicon quantum dots (SiQDs) whose optical and electronic properties are modified by the quantum confinement. Size-tunable optical bandgap allows for emission spectrally tunable in wide UV-to-IR region (260--1100 nm), attractive for the currently developing market of QD-based LEDs and displays \cite{Anikeeva09, Erdem11, Maier-Flaig13}. Non-toxicity \cite{Alsharif09, Bhattacharjee13, Liu13}, bio-degradability \cite{Park09} and superior photo- and pH-stability \cite{He11} of SiQDs open opportunities in traditionally high health risk areas such as medicine or cosmetics. Even though quantum confinement in SiQDs leads to considerable improvement in radiative rates, their magnitude remains still very low (e.g. 10$^4$ s$^{-1}$ for $\sim$2.5 nm H-capped SiQD), compared to direct bandgap materials utilized for light emitter and laser applications. The reason is persistently indirect character of the ``fuzzy" band structure \cite{Hapala13}, rendering radiative transitions to be phonon-assisted. Interestingly, bright fast emission with high rates of $10^7$--$10^9~{\rm s}^{-1}$ has been experimentally observed from various chemically synthesized SiQDs with organic capping \cite{Wilcoxon99, English02, Zou04, Warner05, RossoVasic08, Kusova10, Dohnalova12, Dohnalova13}. The enhancement of the radiative rate in such samples has been partially explained only recently~\cite{Dohnalova13, Kusova14} as related to an appearance of electron states in the $\Gamma$-valley of the Brillouin zone at the SiQD band-edge energies. Hence, the radiative transitions become direct in $k$-space (do not require phonons) and their rates are greatly enhanced. While the tight-binding calculations of Ref.~\cite{Dohnalova13} indicate that the carbon at the QD surface may lead to appearance of the electron states in the $\Gamma$-valley in the absence of strain, Refs.~\cite{Hapala13, Kusova14} suggest that this effect can only rise as a result  of a relatively large ($\sim 3 \%$) ligand-induced strain~\cite{Kusova14}. Hence, origin of the fast emission in the particular case of carbon capping is not completely clear.

In general, the physics of emission from SiQD is challenging and not yet fully controlled due to the complex interplay between the core and surface chemistry and approaches to improve the emission from the SiQDs via surface engineering are understood even less (for current review see, e.g., Ref. \cite{Dohnalova14}). One of the least explored possibilities is the effect of electronegative surface capping. It has been shown that electronegative ligands strongly influence the emission from ultrasmall semi-ionic QDs from materials like CdSe \cite{Schreuder09}. Moreover, the simulations of ultrasmall Si clusters capped with highly electronegative F and Cl \cite{Ramos12, Wang13, Ma11, Ma12} have revealed dramatic changes in HOMO-LUMO energies. Here, we explore general influence of the electronegative capping/ environment on the carrier wavefunctions and radiative transitions in larger SiQDs of sizes $\sim$1.8--4~nm. We present a qualitative explanation of the origin of the $\Gamma$-valley states at the bottom of the conduction band reported in Refs.~\cite{Dohnalova13, Hapala13, Kusova14} and draw important conclusions for the enhancement of the emission from SiQDs material and possibility to achieve direct bandgap SiQDs.

\begin{figure*}
\centering
\includegraphics[width=\textwidth]{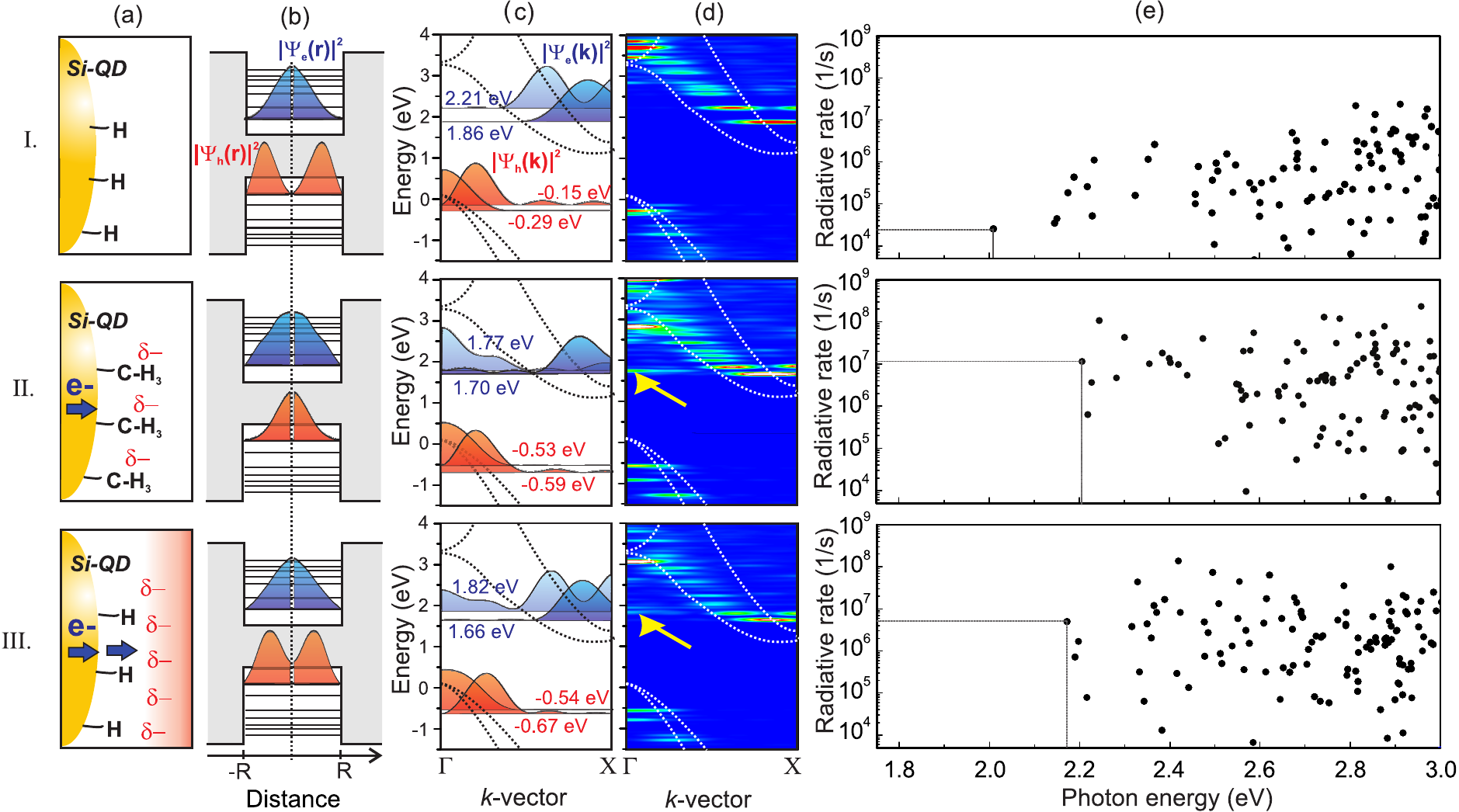}
\caption{Influence of local electric field induced by electronegative capping/ environment in three different $\sim$2.5 nm SiQDs systems: (I) H-capped SiQDs; (II) SiQDs capped with covalently bonded electronegative ligand (here C-capped SiQDs); (III) H-capped SiQDs in electronegative environment. Separate columns depict: (a) Sketch of the system; (b) Electron and hole wavefunction density of ground state (HOMO-LUMO) in real space; (c) Wave-vector resolved density $|\Psi_{\bm k}|^{2}$ of the two lowest electron states (LUMO1, LUMO2) and hole states (HOMO1, HOMO2); corresponding energies with respect to valence band maxima are indicated; (d) Wave-vector-resolved density of states (``fuzzy" band structure; the $\Gamma$-valley levels are indicated by yellow arrows); (e) Radiative rates as a function of e-h pair recombination energy with indicated lowest HOMO-LUMO transition.} \label{fig1}
\end{figure*}

In our simulation, we consider 1.8--4 nm SiQDs using $sp^{3}d^{5}s^{*}$ empirical tight-binding (TB) approach~\cite{Jancu98}. Optical transitions are calculated in the diagonal approximation for the coordinate matrix element \cite{Cruz99}. In Figs.~\ref{fig1} are shown overviews of the main results from three simulated systems: H-capped SiQDs as a reference system (panels I), SiQDs with covalently bonded electronegative capping (panels II) and H-capped SiQDs embedded in electronegative environment (panels III). Reference H-capped SiQDs are calculated using the hydrogen parameters from Ref.~\cite{Hill96}. More details on simulation are given in Refs~\cite{Poddubny10,Dohnalova13} and Supplementary Materials. Calculated wave vector-resolved carrier density of states (name ``fuzzy band structure" has been suggested in Ref.~\cite{Hapala13}) in H-capped SiQDs follows the bulk Si dispersion (dashed lines in Fig.\ref{fig1}-Id and Fig.~\ref{fig3}a), in agreement with the DFT simulations in Ref. \cite{Hapala13}. Consequently, the overlap between electron and hole wave function densities in $k$-space remains low (Fig.~\ref{fig1}-Ic,Id), resulting in low phonon-less radiative recombination (Fig.~\ref{fig1}-Ie) and the optical transitions benefit from the assistance of phonons \cite{Moskalenko07}.

\textbf{Electronegative capping.} The concept of electronegativity has been revised many times to account for the effects of the environment and the difference between molecules and solid states systems \cite{Allen89}. For our purpose, electronegativity $\chi_{spec}$ is introduced as a weighted average of $s$- and $p$-orbital energies ($\epsilon_p$ and $\epsilon_s$, respectively) via equation $\chi_{spec}=(m\epsilon_p+n\epsilon_s)/(m+n)$, where $m$ and $n$ are number of electrons in the $p$- and $s$-orbital, respectively \cite{Allen89}. Consequently, the electronegative ligand -X is simulated by Si-like atom with valence $p$-orbital energies lowered with respect to the bulk Si by the variable $\Delta_{p}$. In general both valence $s$- and $p$-orbital energies need to be shifted, but we found that the overall influence of the $s$-orbital energy shift on presented results is negligible. Shift in $p$-orbital energies of surface atoms results in the ``pull"-effect on the core electronic density, related to the understanding of electronegativity in chemical sciences as the tendency of atom/ligand to attract electrons towards itself \cite{Piela13}. This definition provides straightforward connection to the TB simulations and allows interpretation of the following results in terms of the~``ideal" capping element for enhanced emission in SiQDs system.

\begin{figure}
\includegraphics[width=9cm]{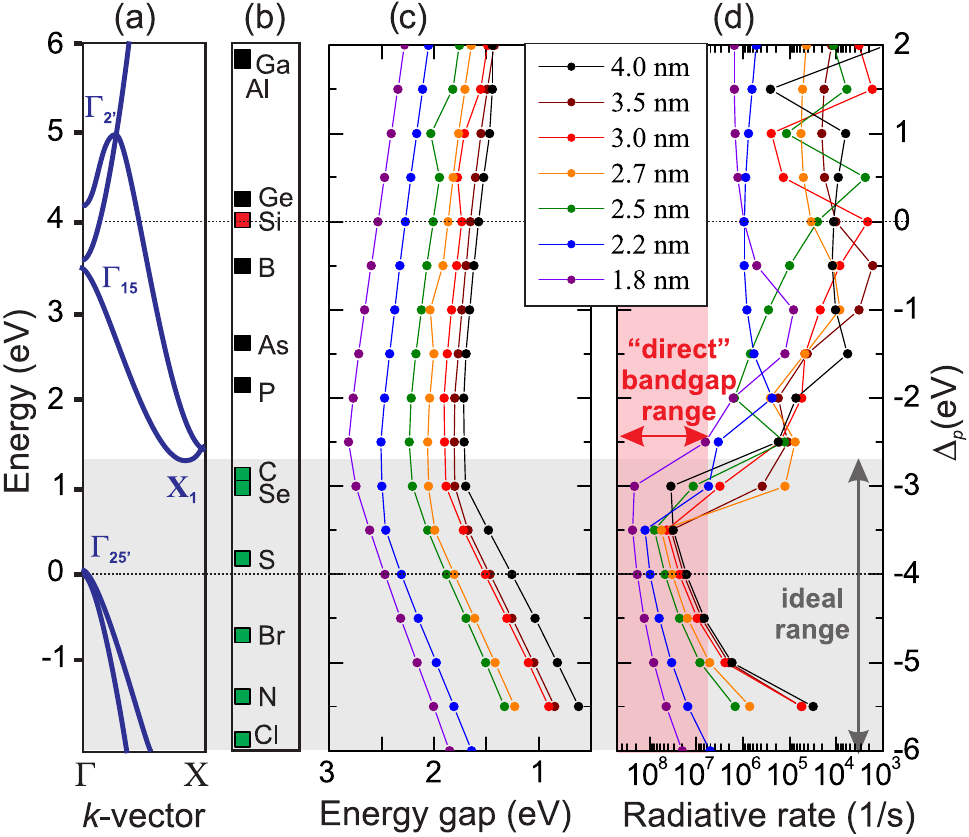}
\caption{Influence of the covalently bonded electronegative capping. (a) Bulk Si band structure. (b) $p$-orbital energy of various elements with respect to silicon (red)(y-axis as in c,d); in green are depicted those elements that are potentially leading to enhanced radiative rate (``ideal range" gray area in panel (d)). (c) Optical bandgap and (d) radiative rates of various sizes of SiQDs (1.8--4 nm) ``capped" with electronegative element with $p$-orbital energy shifted from the bulk Si value by $\Delta_{p}$.}
\label{fig2}
\end{figure}

Fig.~\ref{fig2} summarizes main calculated results for the influence of the covalently bonded electronegative ligand with $p$-orbital energy of the surface atoms  shifted by $\Delta_p$. Panels (a,b) show $p$-orbital energy of various elements with respect to bulk Si band structure, and panels (c,d) show dependence of the band-gap energy and radiative rates, respectively, on $\Delta_p$. Both electron and hole energies decrease for negative $\Delta_{p}$, following the decrease of the surface orbital energy (see Supplementary Materials, Fig. S1e). Result is a non-monotonous energy gap dependence peaking at $\Delta_{p}\approx -2~$eV and rapidly decreasing afterwards (Fig.~\ref{fig2}c). The narrowing of the band gap for strong electronegative capping is consistent with the results obtained for ultra-small Si clusters covered with F and Cl in Refs.~\cite{Ramos12, Wang13, Ma11, Ma12}. Our central result is the strong modification of the phonon-less radiative rate with a prominent maximum for $-6~{\rm eV} < \Delta_{p}<-2.7~{\rm eV}$ (Fig.~\ref{fig2}d, gray area), present for all the studied SiQD sizes 1.8 -- 4 nm. For SiQDs up to $\sim$2.5 nm, the highest radiative rate reaches values of $\sim 10^8~{\rm s}^{-1}$, typical for direct band gap QDs \cite{vanDriel05} (Fig.~\ref{fig2}d, pink area). This defines an ``ideal" range of $\Delta_p$ for enhanced radiative rate, which includes various real elements, such as C, Se, S, Br, N and Cl (green symbols in Fig.~\ref{fig2}b). From these, C is the material most frequently used in experimental studies. $\Delta_{p}=-3$~eV well reproduces tight-binding parameters of carbon in literature \cite{Laref08} and allows us to effectively simulate capping by -CH$_3$ methyl ligands (``C-capped SiQDs").

To provide further elucidation into the origin of the fast radiative rates, we compare results in Fig.~\ref{fig1} for $\sim$2.5 nm H- and C-capped SiQDs simulated with $\Delta_{p}=0$ and $\Delta_{p}=-3$~eV, respectively. The ``pull" effect of the electronegative capping on the electrons towards the SiQD surface can be seen from comparison of the real-space electronic densities in H- and C-capped SiQDs (Fig.~\ref{fig1}-Ib, IIb). Also the structure of the hole eigenstates has changed: In H-capped SiQDs the ground hole state wavefunction is of $p$-type character, which reflects in small amplitudes in the coordinates origin both in real- (Ib) and $k$-space (Ic) and the first excited hole state is of $s$-type; Capping with carbon decreases the energy of the $p$-type state, and the $s$-type hole state becomes the ground one (IIb, IIc) (for more details see Supplementary Materials, Figs. S6, S7). This shows as additional maxima in the $\Gamma$-valley in the wave-vector resolved electronic density (``fuzzy band-structure" \cite{Hapala13}), indicated by yellow arrow in Fig.~\ref{fig1}-IId, well resolved already for the ground states of electrons (panel IIc). This effect is size-dependent, as shown in Fig.~\ref{fig3} for both H- and C-capped SiQDs (for more details see Supplementary Materials, Figs. S2, S3).

Occurrence of the new $\Gamma$-valley levels allows for fast $\Gamma-\Gamma$ phonon-less radiative recombination with holes, leading to the enhanced radiative rate (Fig.~\ref{fig1}-IIe), observed experimentally (e.g. Refs. \cite{Kusova10, Dohnalova12, Dohnalova13}). As a result, the SiQDs effectively behave as if made from the direct band-gap semiconductor. Interestingly, the highest radiative rates found in Fig.~\ref{fig2}d correspond to the $p$-orbital energy shift $\Delta_{p}=-3.5~$eV, when the $p$-orbital level is positioned within the Si bandgap (Fig.~\ref{fig2}a). This could indicate that the strong modification of the confined carrier states is a result of resonant enhancement of the mixing between the electron states in the $X$-valley and the hole states in the $\Gamma$-valley. While the valley mixing, including the $\Gamma-X$ mixing, is well known in semiconductor nanostructures \cite{Poddubny12, Friesen10, Bulutay07, Zunger06, Jancu04, Ivchenko93}, in our case the effect is much stronger. The resonant effect for the $p$-orbital might also explain why the radiative rate and band gap energies in Figs.~\ref{fig2}c,d are independent of $s$-orbital energies and only relatively weakly dependent on the $\Delta_p$ for $p$-orbital energy falling into the bulk Si band-gap energies.

\begin{figure}
\includegraphics[width=8.5cm]{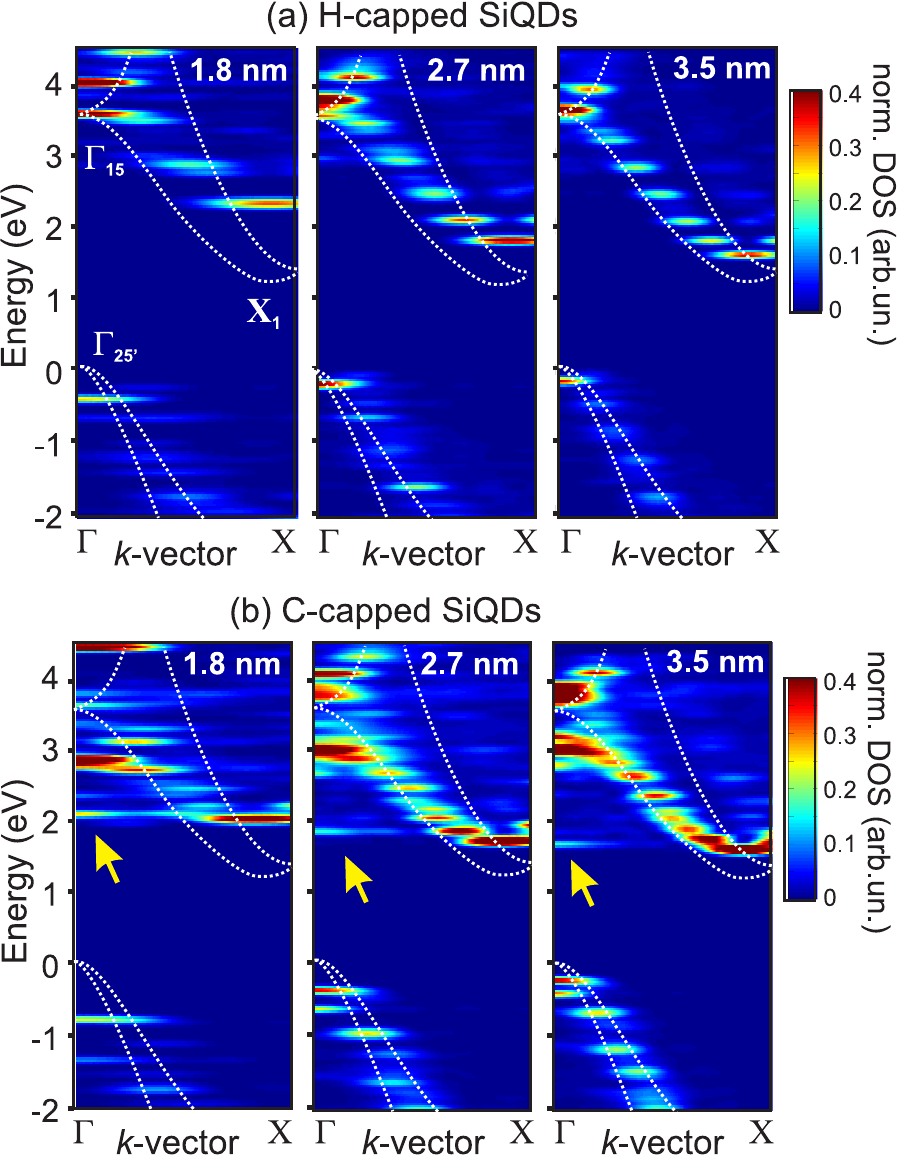}
\caption{Wave vector-resolved density of carrier states (``fuzzy band structure") in (a) H-capped SiQD and (b) C-capped SiQDs for the sizes 1.8, 2.7 and 3.5 nm. White dotted lines represent bulk Si band structure. Density of states is normalized to the high density around 3--3.5 eV, i.e. to the density of states of former bulk $\Gamma_{15}$ band. Yellow arrows indicate position of the newly formed $\Gamma$-valley levels.}\label{fig3}
\end{figure}

\textbf{Electronegative environment.} In the system with covalently bonded electronegative capping, two effects seem to be responsible for the enhanced rate -- (i) the $\Gamma-X$ valley mixing induced by the covalently bonded ligand and (ii) the ``pull" on the electronic density by the effective local electric-field induced by the electronegative ligand. In order to elucidate the role of the second effect, it is instructive to consider another model system -- H-capped SiQD embedded in an electronegative environment, where the electronegative element is not directly covalently bonded (sketch in Fig.~\ref{fig1}-IIIa). In our TB model, effect of electronegative environment is simulated phenomenologically by the gradual change in $p$-orbital energies of Si atoms through the whole SiQD, following equation $E_p(R)=E_p({\rm Si})+U(R/R_{\rm max})^2$, where $E_p(R)$ is energy of the $p$-orbital level for Si atom on position $R$, $E_p({\rm Si})$ is energy of $p$-orbital in bulk Si, $U$ is the variable factor defining magnitude of the electronegative effect (local field) and $R_{\rm max}$ is radius of the SiQD (see also Fig. S4). This effect can be understood in terms of ``band-bending", such as in metal-oxide-semiconductor (MOS) capacitors. For the simulation in Figs.~\ref{fig1}-IIIb--IIIe, the strength of the effective field is chosen such that the  $p$-orbital energy of the surface atoms is lowered by 3~eV ($U=-3$~eV). The ``pull" effect on the electronic density is softer than that introduced by the covalently bonded electronegative ligand (compare Figs.~\ref{fig1}-Ib, IIb and IIIb). Consequently, the newly formed $\Gamma$-valley peaks in the wave vector-resolved density of states are qualitatively very similar to those in the C-capped SiQDs, only less pronounced (Figs.~\ref{fig1}-IIIc, IIId). Resulting radiative rates (Fig.~\ref{fig1}-IIIe) are enhanced slightly less than for C-capped SiQDs of the same size. For more detailed study of the electronegative environment, see Supplementary Materials, Figs. S4--S7.

\textbf{Conclusions.} We conclude that electronegative capping/ environment potentially allows to achieve fast direct radiative transitions as a result of either $\Gamma-X$ valley mixing induced by the covalently bonded ligand and/or the ``pull" on electronic density by the effective local electric-field induced by the electronegative agent. Let us discuss the ideal system. In the case of covalently bonded electronegative capping, from the pool of available elements (B, C, N, O, F, P, S, Cl and Ge), it appears that only those with lower $p$-orbital energy than Si are beneficial, such as B, C, N, O and F (see Fig.~\ref{fig2}b, gray area). While B and C have advantageously positioned $p$-orbital, B acts as a dopant and is therefore not suitable, since non-radiative losses due to Auger recombination can be expected. Furthermore, it has been demonstrated that B tends to be incorporated rather inside the SiQDs than on the surface~\cite{Pi08}. C, on the other hand, has the same valency as Si, higher electronegativity and well positioned $p$-orbital energy inside the optical bandgap of SiQD. N and O have different valency than Si and tend to form double or bridging bonds. In case of O, double bonded species \cite{Puzder02, Puzder02PRL} and bridging Si-O-Si bond \cite{Vasiliev02} lead to deep interband states. Similar effects have been presented also for S \cite{Puzder02PRL}. Formation of defect states has been also recently suggested for N \cite{Dasog13}. For F and Cl cappings in ultrasmall Si clusters, DFT  calculations~\cite{Wang13, Ma11, Ma12} indicate strong impact on the electronic states for higher than 50\% surface coverage. Full coverage of surface by Cl was shown to lead to the dramatic drop in the  valence and conduction band energies with the band gap smaller than in bulk Si \cite{Ma11}. Furthermore, 50\% surface coverage enhances the radiative rate $\sim$40 times  up to $\sim$1.2 10$^8$ s$^{-1}$ \cite{Ma11}, while at larger  coverage the rate decreases. The calculations for F have revealed similar, but less dramatic drop of the valence and conduction band states energy and also a slight increase in the radiative rate  in case of partial surface coverage, followed by a drop for full surface coverage \cite{Ma12}. Our results indicate  that for large SiQDs with $D\gtrsim 4$~nm, covered with F or Cl, one can achieve the band gap smaller than bulk Si and, simultaneously, relatively large radiative rates $\sim 10^{7}~\rm s^{{-1}}$ (Fig.~\ref{fig2}d). However, such material might not be chemically stable. Hence, up till now, C-linked species remain the most beneficial realized ligands for enhanced radiative rate (and enhanced efficiency of emission) of SiQDs. C-linked capping is easily accessible via organic capping with molecules such as alkyl chains, provided by various synthesis techniques. To further improve the electronegativity effect, e.g. teflon chains -(CF$_2$)$_n$-CF$_3$ might offer an ideal capping, taking advantage of both effects ---  the resonant $\Gamma-X$ mixing due to carbon, and the ``pull'' effect caused by enhanced local field by proximity of the highly electronegative fluorine. In summary, electronegative capping is one of the least explored possibilities for tailoring optical properties of nanoscale objects, that opens doors for a variety of fascinating effects, bridging physics, chemistry, and biology through enabling direct bandgap-like SiQDs with greatly enhanced radiative rate.

\section*{Acknowledgements}
ANP acknowledges support from the Russian Foundation for Basic Research and the ``Dynasty'' foundation, KD acknowledges Stichting voor de Technologische Wetenschappen (STW) and Stichting der Fundamenteel Onderzoek der Materie (FOM) funding. Both authors acknowledge fruitful discussions with I. Yassievich, A. Prokofiev, M. Nestoklon, D. Vanmaekelbergh, J. M. J. Paulusse, H. Zuilhof and T. Gregorkiewicz.


\newpage
\newpage
\renewcommand{\thefigure}{S\arabic{figure}} 
\section{Supplementary Materials}
\begin{figure*}
\centering
\includegraphics[width=\textwidth]{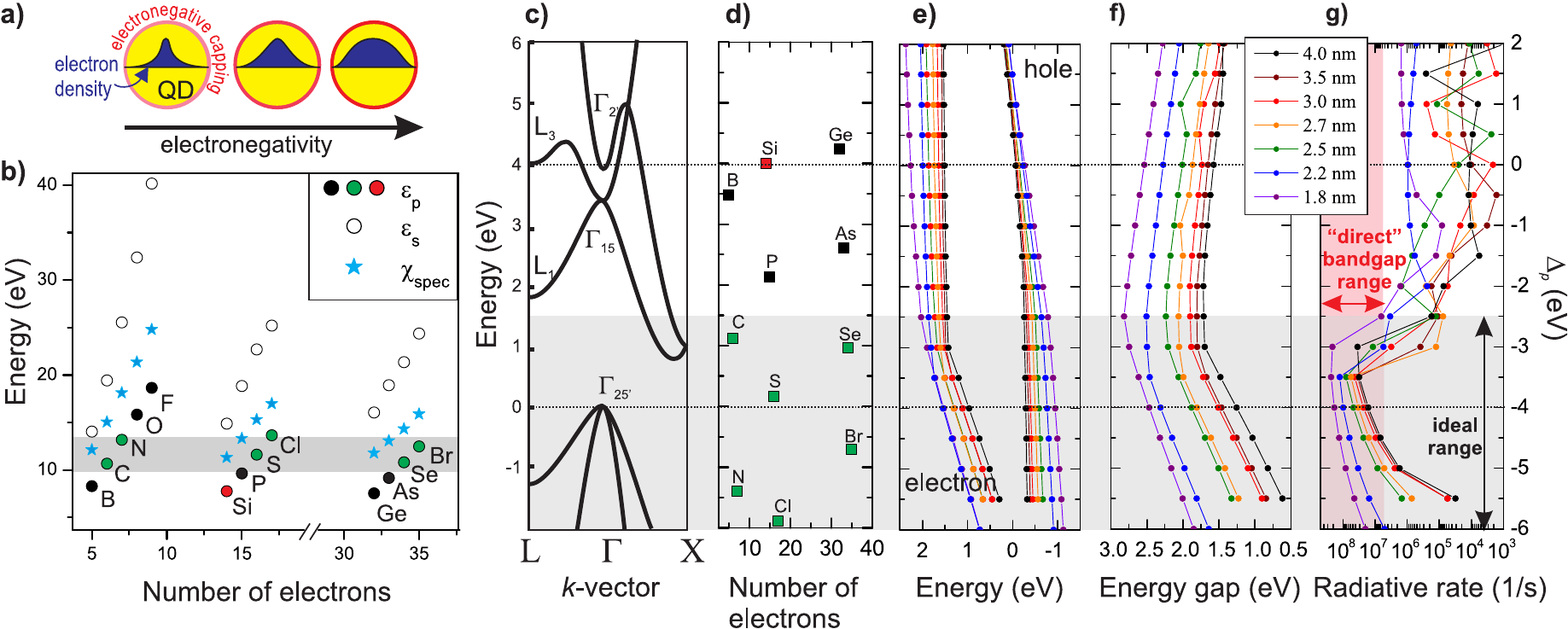}
\caption{\textit{General influence of the covalently bonded electronegative capping. (a) Schematic sketch of the influence of electronegative capping on the electronic density in real-space in QD. (b) $s$- and $p$-orbital energies and electronegativity calculated from Eq. \eqref{eq1} for elements with higher electronegativity than silicon. (c) Bulk Si band structure (energy is set as zero at the top of valence band). (d) Relative energy position of $p$-orbital of various elements with respect to silicon (red)(y-axis is common for d-g); in green are depicted "ideal" capping elements that are potentially leading to enhanced radiative rate (gray area in (g)). (e) Electron (LUMO) and hole (HOMO) energies, (f) band gap energies (HOMO-LUMO transition) and (g) radiative rates for various sizes of SiQDs (1.8--4 nm) ``capped" with the virtual element X (with shifted $p$-orbital energy by $\Delta_{p}$ from the bulk Si value $E_{p}({\rm Si})$).}} \label{figS1}
\end{figure*}

\textbf{Electronegative capping.} In chemical bond, the more electronegative atom/ligand tends to attract electrons towards itself \cite{Piela13} in pursuit of lower total energy of the system. Physically most insightful definition of electronegativity $\chi_{spec}$ for our system is therefore a weighted average of $s$- and $p$-orbital energies ($\epsilon_p$ and $\epsilon_s$, respectively) \cite{Allen89} by equation
\begin{equation}\label{eq1}
\chi_{spec}=(m\epsilon_p+n\epsilon_s)/(m+n),
\end{equation}
where $m$ and $n$ are number of electrons in the $p$- and $s$-orbital, respectively. In QDs capped with ligand that is more electronegative than the constituent atoms of the QD, electronegative capping  ``pulls" the electronic density towards the surface in real-space (see schematic sketch in Fig.~\ref{figS1}a) and consequently through Fourier transform also affects it in $k$-space. Possible capping elements for SiQD for this purpose are shown in Fig.~\ref{figS1}b. The above definition of the electronegativity (Eq.~\eqref{eq1}) allows us to model electronegative ligand X as a virtual atom with valence $p$-orbital energies lowered with respect to the bulk Si by a variable $|\Delta_{p}|$. In a general case, both $s$- and $p$-orbital energies should be shifted. However, we found that shift of $s$-orbital energy has only negligible influence on final results, so we will only modify the $p$-orbital energy. Figs.~\ref{figS1}c,d show the bulk Si band structure and $p$-orbital energies of various atoms with higher electronegativity than Si that are considered for capping. The energy axes in Figs.~\ref{figS1}d--g correspond to the energy shift $\Delta_{p}$ from the bulk Si $p$-orbital energy value $E_{p}({\rm Si})$. Figure~\ref{figS1}e shows electron (LUMO) and hole (HOMO) level energies for different SiQD sizes between 1.8 and 4 nm. For small negative values of $\Delta_{p}$, the electron ground  energy is only weakly affected while the hole energy decreases faster. For large negative $\Delta_{p}$, corresponding to the strong surface electronegativity, the electron energy decreases fast, while the hole energy saturates. This leads to non-monotonous energy gap dependence on $\Delta_{p}$ (Fig.~\ref{figS1}f), peaking at $\Delta_{p}\approx -2~$eV and rapidly decreasing afterwards. For complete picture we show again also Fig.~\ref{figS1}g, where radiative rates for the electron-hole pair ground state (HOMO-LUMO) transition are indicated. As discussed in the manuscript, radiative rates are dramatically enhanced for surface $p$-orbital energy lowered by more than 2.5 eV. This range is depicted by the gray area in Figs.~\ref{figS1}b-g. The pink area depicts the``direct bandgap range" zone, where the radiative rates are enhanced above $\sim$10$^7$ s$^{-1}$. Such high radiative rates are typical for direct bandgap materials, such as CdSe QDs. The elements that have $p$-orbital energy in this region are colored in green in Figs.~\ref{figS1}b,d and are considered as ideal capping. One of them is carbon, which can be found in commonly used organic ligands (e.g. alkyl chains).\\

\begin{figure*}
\centering
\includegraphics[width=14cm]{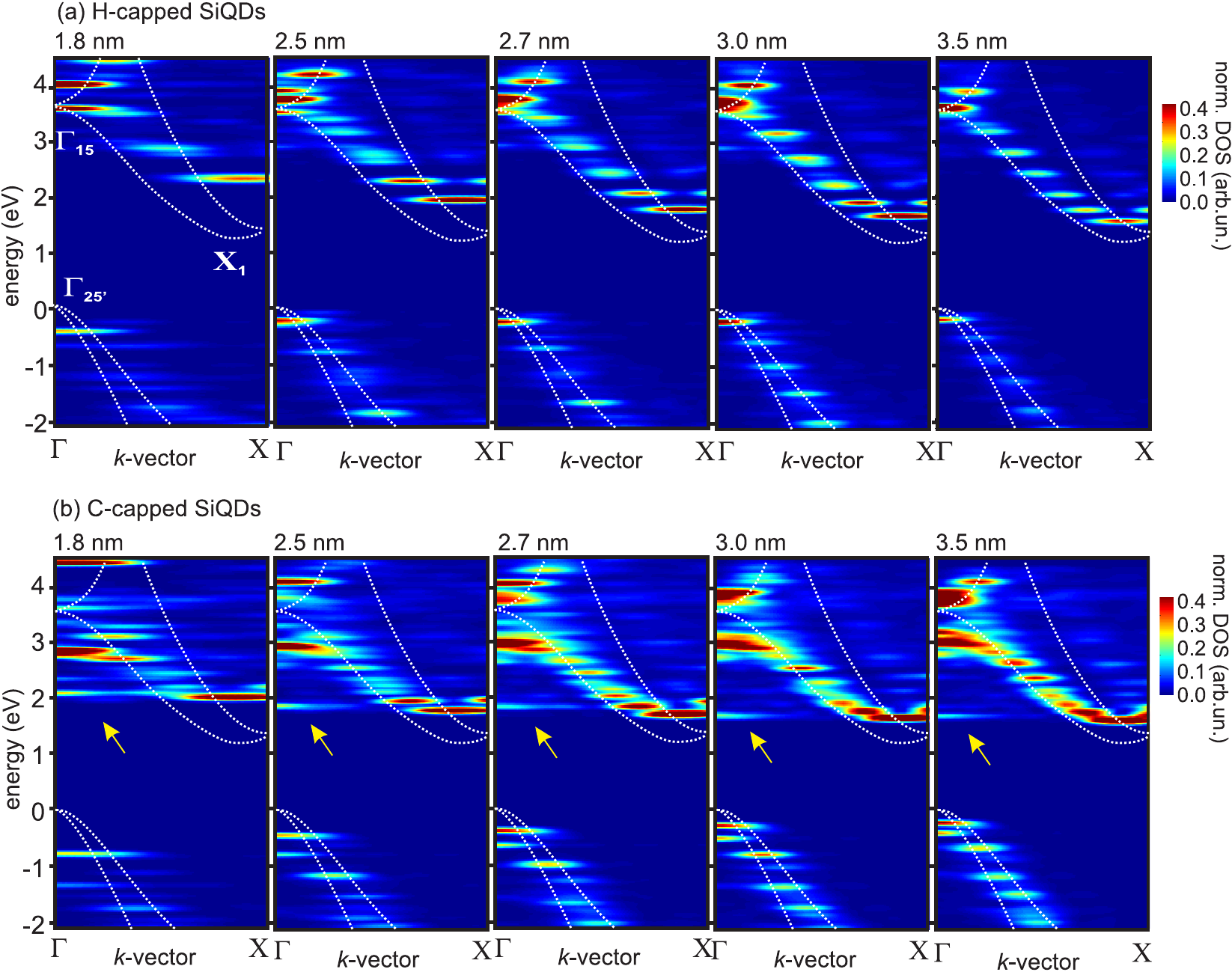}
\caption{\textit{Wave vector-resolved density of carrier states in (a) H-capped SiQDs and (b) C-capped SiQDs for different SiQD sizes between 1.8 and 3.5 nm; Density of states is normalized to the high density of states around 3--3.5 eV (former bulk $\Gamma_{15}$ band).}} \label{figS2}
\end{figure*}

\begin{figure*}
\centering
\includegraphics[width=14cm]{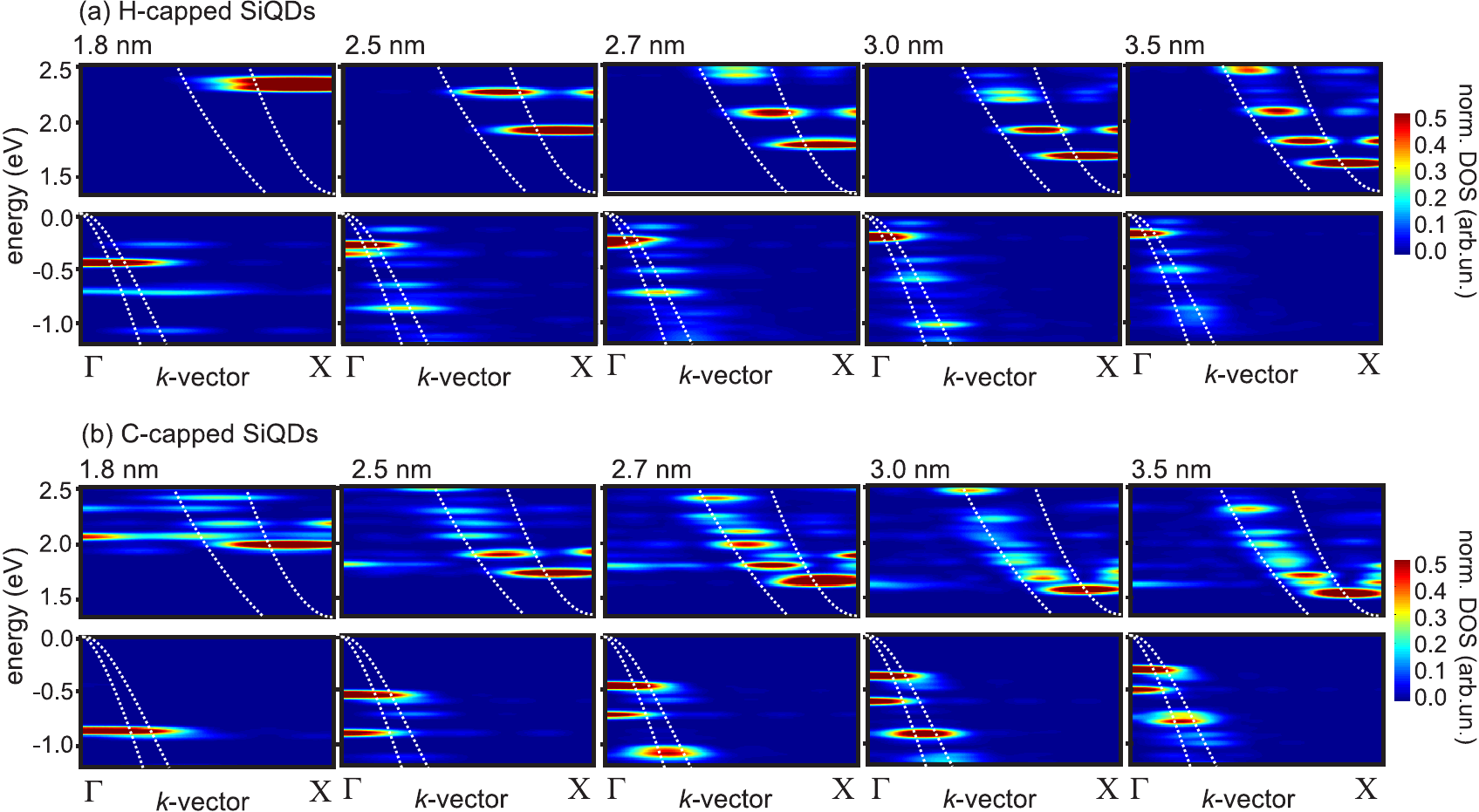}
\caption{\textit{Detail of the graphs from Fig.~\ref{figS2} close to the top of valence band and bottom of conduction band.}} \label{figS3}
\end{figure*}

Figs.~\ref{figS2} and \ref{figS3} show the bulk Si band structure (white lines) compared with the wave vector-resolved electronic density (appropriate name ``fuzzy" band structure appeared in Ref. \cite{Hapala13}) of SiQDs of sizes between 1.8 nm and 3.5 nm capped with (a) hydrogen (calculated with $\Delta_{p}=0$) and (b) covalently bonded electronegative capping ($\Delta_{p}=-3$~eV). The properties of the latter are very similar to those of carbon-linked methyl (-CH$_3$) groups ("C-capped SiQDs"). Energy levels in the H-capped SiQDs follow the shape of the indirect bulk Si band structure (white lines). The C-capped SiQDs, on the other hand, show new $\Gamma$-valley levels at the bottom of the conduction band. Major changes appear also for the higher energy levels close to the former bulk $\Gamma_{15}$ band. Figure~\ref{figS3} shows the details of Fig.~\ref{figS2} close to the bottom of the conduction band (LUMO) and the top of the valence band (HOMO).

\begin{figure*}
\centering
\includegraphics[width=14cm]{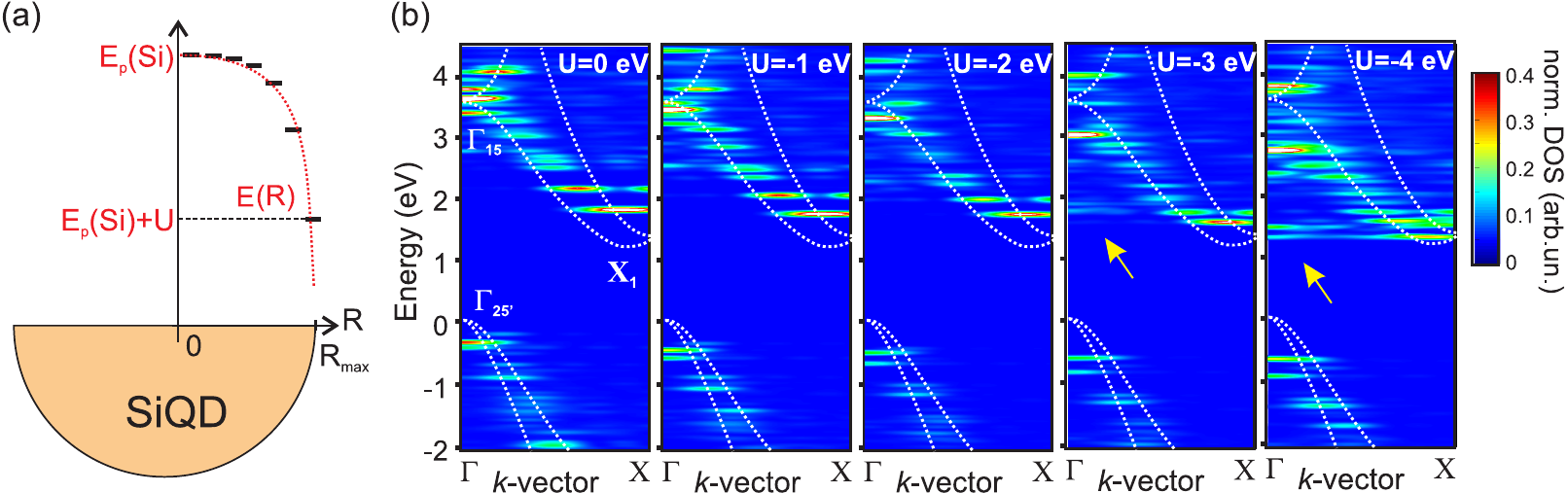}
\caption{\textit{(a) Sketch of the band bending. (b) Wave vector-resolved density of carrier states in $\sim$2.5 nm SiQD capped with hydrogen and immersed in electronegative environment, simulated by band bending using Eq.~\eqref{eq2} for different strengths of the local field. Density of states is normalized to the high density of states around 3-3.5 eV (former bulk $\Gamma_{15}$ band).}} \label{figS4}
\end{figure*}

\begin{figure}
\centering
\includegraphics[width=0.45\textwidth]{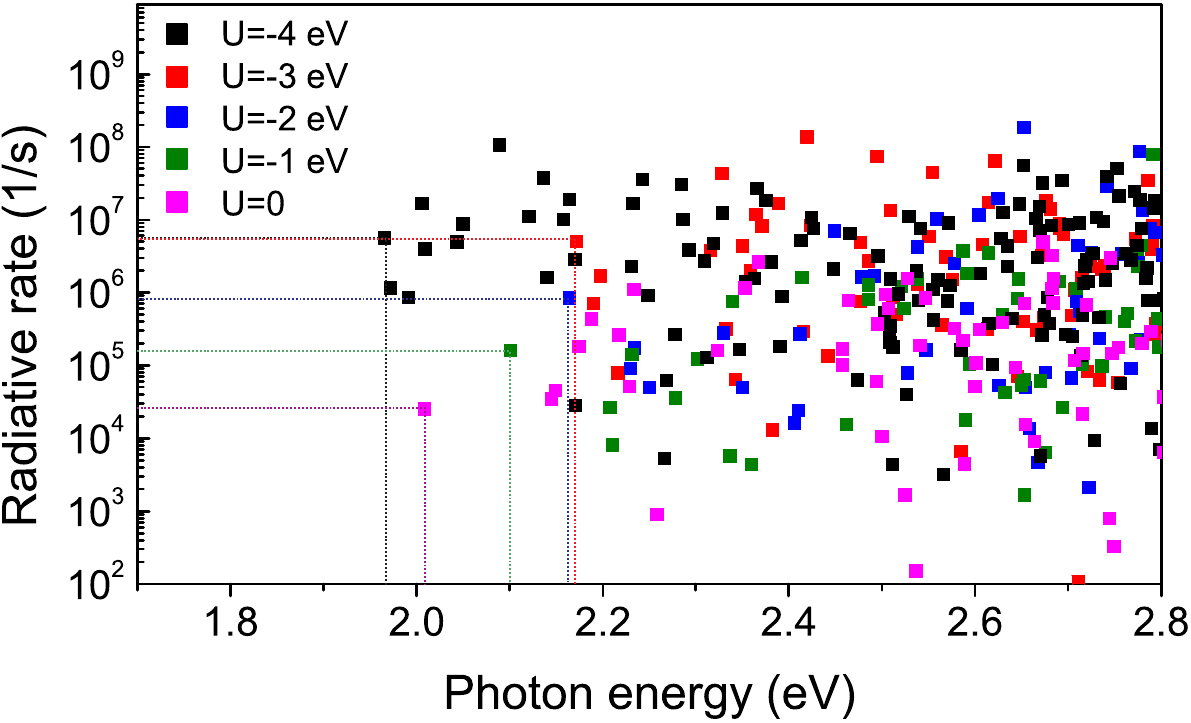}
\caption{\textit{Radiative rates for e-h pair recombination in H-capped $\sim$2.5 nm SiQDs embedded in electronegative environment of different strengths.}} \label{figS5}
\end{figure}

\textbf{Electronegative environment.} To separate influence of the covalent bonding and local field created by the electronegative bond, we phenomenologically simulate the system of H-capped SiQDs immersed in electronegative environment. This is achieved by gradual ``band-bending" (very much like that e.g. in metal-oxide-semiconductor (MOS) structures; see the sketch in Fig.~\ref{figS4}a), where energy level of the valence $p$-orbital of Si atoms is gradually lowered towards the surface, following equation
\begin{equation}\label{eq2}
E_p(R)=E_p({\rm Si})+U\left(\frac{R}{R_{\rm max}}\right)^2,
\end{equation}
where $E_p(R)$ is energy of $p$-orbital level for Si atom on position $R$, $E_p({\rm Si})$ is energy of $p$-orbital in bulk Si, $U$ is variable factor defining magnitude of the electronegative effect (local field) and $R_{max}$ is radius of the SiQD (Fig.~\ref{figS4}a). Figure~\ref{figS4}b shows the resulting wave vector resolved density of carrier states in such SiQD system of size $\sim$2.5 nm for various electronegativity strengths. The value $U=0$ eV corresponds to typical H-capped SiQDs and $U=-3$ eV could be directly compared to the previous case of covalently bonded methyl-like capping, where $\Delta_p= -3$~eV. Apparently, covalent bonding is more efficient, since the $\Gamma$-valley levels are more pronounced for the $\sim$2.5 nm methyl-capped SiQD in Figs.~\ref{figS2}, \ref{figS3}. In Fig.~\ref{figS5} are shown the resulting radiative rates depending on  the e-h pair recombination energies for the same bending parameters. We can see that if we apply higher electronegativity ``pull", the radiative rate is higher as a result of more pronounced $\Gamma$-valley states at the bottom of the conduction band, which is in qualitative agreement with the results from the previous system of covalently bonded electronegative capping.

\begin{figure}[b!]
\centering
\includegraphics[width=0.45\textwidth]{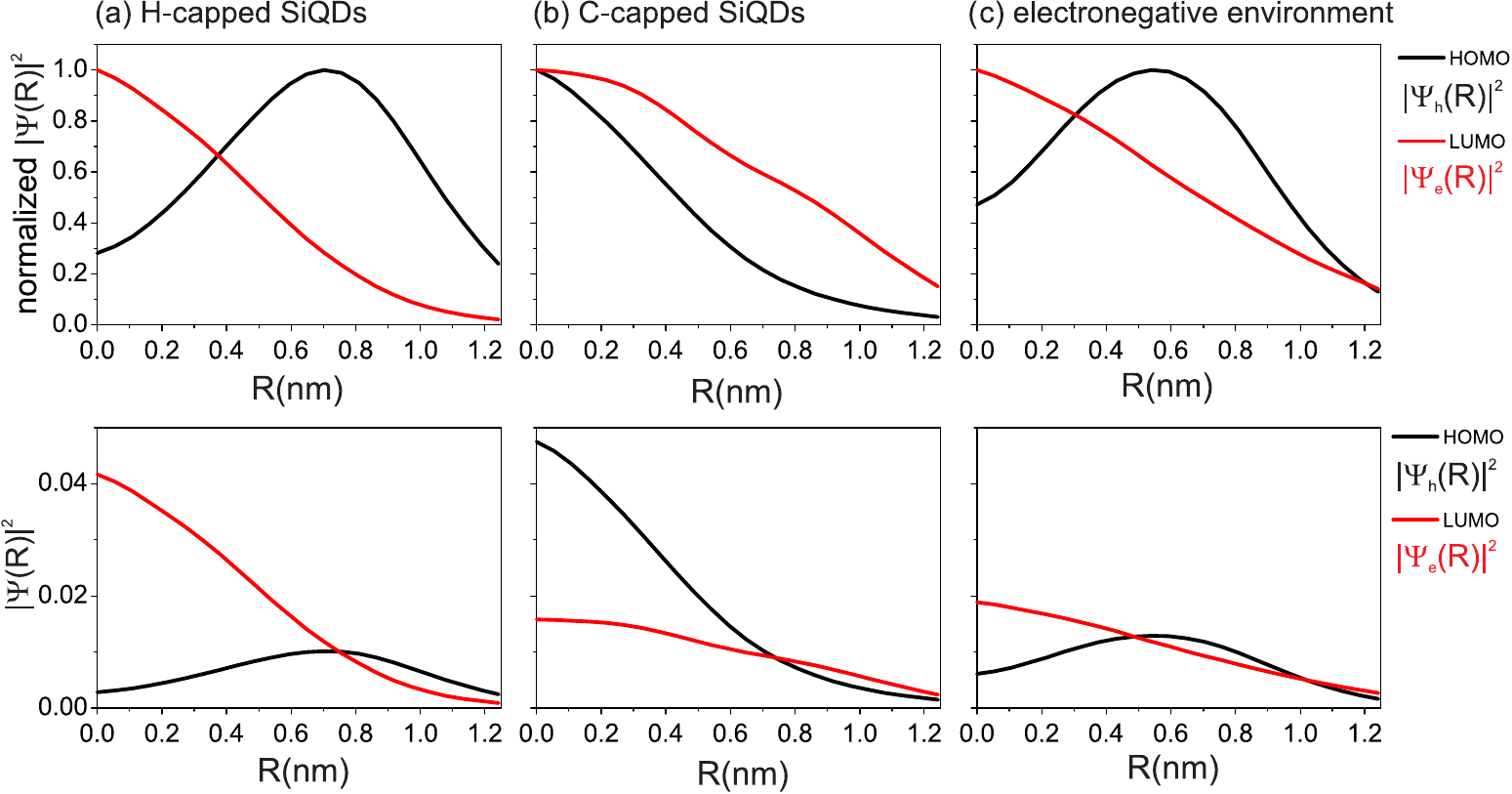}
\caption{\textit{Electron and hole lowest excited state (HOMO-LUMO) wavefunction density $|\Psi(r)|^2$ in real-space for $\sim$2.5 nm (a) H-capped SiQDs; (b) C-capped SiQDs, simulated by $\Delta_p=-3$ eV; (c) H-capped SiQDs immersed in electronegative environment, simulated by $U= -3$~eV. Upper panels show normalized $|\Psi(r)|^2$, lower panels show the original ``as-calculated" data.}} \label{figS6}
\end{figure}

\begin{figure}[b!]
\centering
\includegraphics[width=0.45\textwidth]{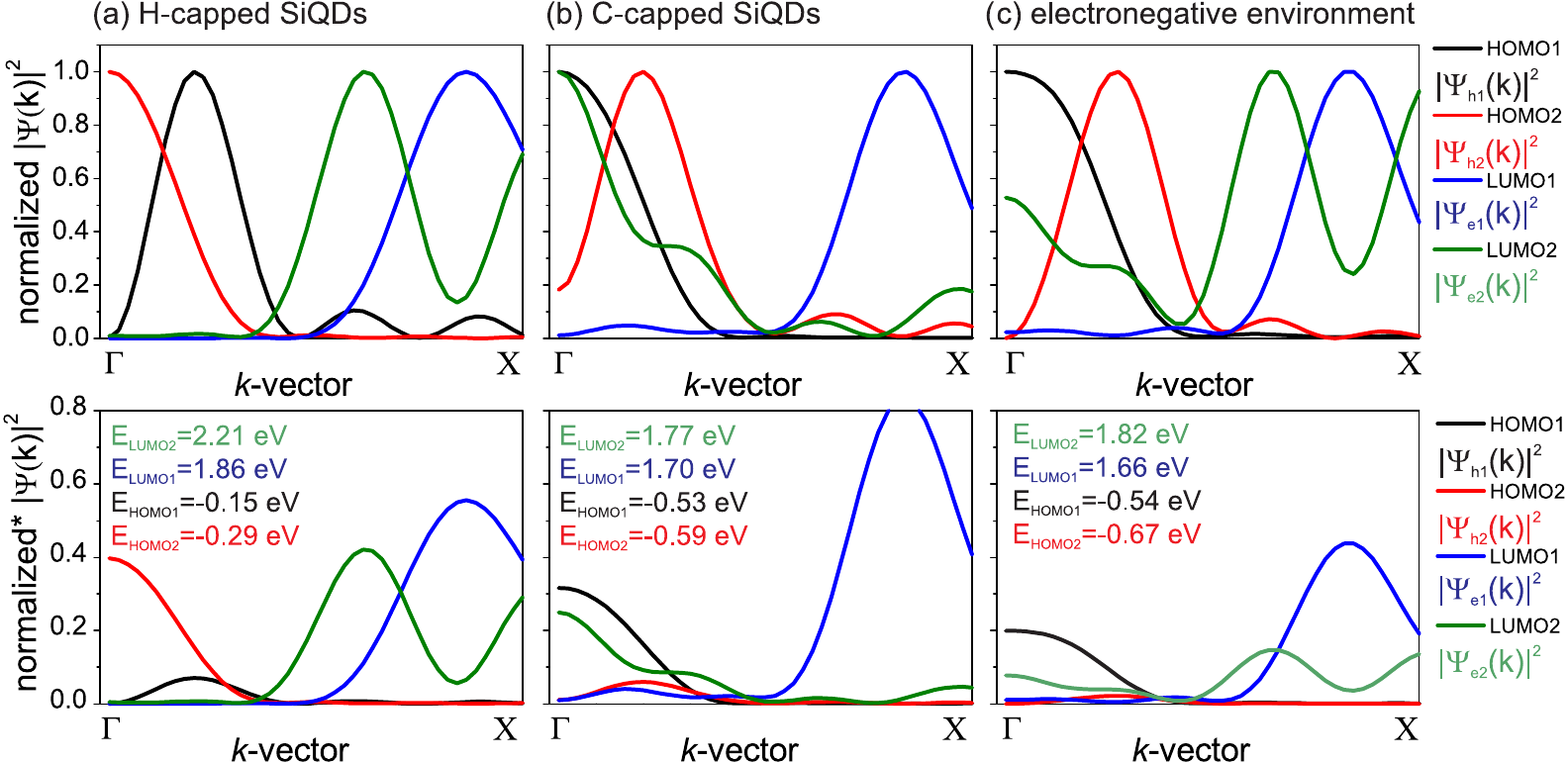}
\caption{\textit{Wave function density for the two lowest electron and hole excited states (HOMO1, HOMO2 and LUMO1, LUMO2) in $k$-space ($|\Psi(\bm k)|^2$) for $\sim$2.5 nm (a) H-capped SiQDs; (b) C-capped SiQDs, simulated by $\Delta_p = -3$~eV; (c) H-capped SiQDs immersed in electronegative environment, simulated by $U= -3$~eV. Upper panels show normalized $|\Psi(\bm k)|^2$, lower panels show $|\Psi(\bm k)|^2$ normalized to the high density of states around 3--3.5 eV, i.e. to the density of states of former bulk $\Gamma_{15}$ band.}} \label{figS7}
\end{figure}

\textbf{Comparison of the electronic densities in real- and $k$-space.} To better understand the direct bandgap-like transitions in both studied systems --- C-capped SiQDs of size $2.5$ nm (simulated with $\Delta_p=-3$~eV) and H-capped SiQD of the same size in electronegative environment (simulated with bending $U$=-3 eV), we plot the electron (LUMO) and hole (HOMO) wavefunction densities in both real-space (Figs.~\ref{figS6}) and $k$-space (\ref{figS7}). Particularly, we present the real space density  $|\Psi(r)^{2}|$ obtained by adding the absolute values of the tight-binding expansion coefficients $|C_{\nu}(\bm a)^{2}|$ for different $s,p,d,s^{*}$ orbitals $\nu$ ($\bm a$ is the atom coordinate), convoluting in real space with a Gaussian with  the dispersion $\sim 0.3$~nm and averaging over the angle of the vector $\bm r$. The $k$-space density of states is calculated in a similar way. For each confined state $|n\rangle$ with the energy $E_{n}$ we find the Fourier component $C_{\nu(\bm k)}$ and sum the squared absolute values of the Fourier components for different orbitals, $\rho_{n}(\bm k)=\sum_{\nu}|C_{\nu}(\bm k)|^{2}$. For Fig.~2d in the main text, the probability $\rho_{n}(\bm k)$ is also averaged over  equivalent $\Delta$ directions of the vector $\bm k$. To calculate the total wave-vector-resolved density of states  we multiply $\rho_{n}(\bm k)$ by the Gaussian $\exp[-(E-E_{n})^{2}/\delta E^{2}]$ with $\delta E=30~$meV and sum over all the states $|n\rangle$.

From comparison of Figs.~\ref{figS6}a,\ref{figS6}b and \ref{figS6}c (red lines), the real-space ``pull" effect on the electronic density $|\Psi_e(r)|^2$ by the electronegative cappimg/environment can be seen. The electronegative capping also changes structure of the hole eigenstates $|\Psi_h(r)|^2$ --- in the H-capped QDs the ground hole state wavefunction is of $p$-type character (black lines in Figs.~\ref{figS6}a). C-capping decreases the energy of the $p$-type state, and the $s$-type hole state becomes the ground one (black line in Figs.~\ref{figS6}b). H-capped SiQD in electronegative environment shows similar type of wavefunction densities as typical H-capped SiQDs, only with small modifications (compare Figs.~\ref{figS6}a and \ref{figS6}c). Wavefunction densities in $k$-space are shown for the two lowest excited states for electron (LUMO1, LUMO2) and hole (HOMO1, HOMO2) in Figs.~\ref{figS7}a,b and c for the three simulated systems. In upper panels the  normalized Fourier components $|\Psi(\bf k)|^2$ are shown.  The lower panels present the density of states normalized to the high density around 3--3.5 eV, i.e. to the density of states of former bulk $\Gamma_{15}$ band. In H-capped SiQDs, lowest states of electron and hole do not overlap much, which results into typical low radiative rates ($\sim10^4~{\rm s}^{-1}$ for the lowest three transition energies). In the C-capped SiQD (Fig.~\ref{figS7}b), the two lowest electronic states are separated by $\sim$70 meV and hole states about $\sim$60 meV. The $k$-profiles of the density of states are dramatically modified, when compared to those in H-capped SiQD --- the $s$-like state of the hole becomes the energetically lowest one and electron levels have increased contributions in the $\Gamma$-valley. This leads to the observed fast radiative rates, enhanced by the factor 10$^3$, when compared to the H-capped SiQDs ($U=0$ in Fig.~\ref{figS5}). Interestingly, similar $k$-space profile occurs also in the H-capped SiQDs embedded in an electronegative environment (Fig.~\ref{figS7}c). Resulting enhancement of the radiative rate is slightly less than for covalently bonded electronegative methyl-like species (C-capped SiQDs), but is still about 400 times higher than in H-capped SiQD of the same size (Fig.~\ref{figS5}).

\end{document}